\documentclass[usenatbib]{mn2e}
\usepackage{psfig,morefloats,url}
\usepackage[multiple]{footmisc}

\footnotesize
\newdimen\digitwidth    
\setbox0=\hbox{\rm0}
\digitwidth=\wd0
\catcode`!=\active
\def!{\kern\digitwidth}
\normalsize

\title[The Parkes multibeam pulsar survey: VII.] {The Parkes multibeam
pulsar survey: VII. Timing of four millisecond pulsars and the underlying
spin period distribution of the Galactic millisecond pulsar population}
\author[D.~R.~Lorimer et al.]
{D.~R.~Lorimer$^1$\thanks{Email: Duncan.Lorimer@mail.wvu.edu},
P.~Esposito$^{2,3}$, 
R.~N.~Manchester$^4$, 
A.~Possenti$^5$,
A.~G.~Lyne$^6$, 
\newauthor
M.~A. McLaughlin$^1$, M.~Kramer$^7$, G.~Hobbs$^4$, I.~H.~Stairs$^8$, 
M.~Burgay$^5$, 
\newauthor
R.~P. Eatough$^7$, M.~J.~Keith$^6$, A.~J.~Faulkner$^9$, N.~D'Amico$^{5,10}$, 
F.~Camilo$^{11}$,
\newauthor
A.~Corongiu$^5$ and F.~Crawford$^{12}$
\\
$^1$ Department of Physics and Astronomy, West Virginia University, 
PO~Box~6315, Morgantown, WV~26506, USA\\
$^2$Istituto di Astrofisica Spaziale e Fisica Cosmica - Milano, INAF,
via E. Bassini 15, I-20133 Milano, Italy\\
$^3$Harvard--Smithsonian Center for Astrophysics, 60 Garden Street,
Cambridge, MA 02138, USA\\
$^4$ CSIRO Astronomy and Space Science, PO~Box~76, Epping NSW~1710, Australia\\
$^5$ Osservatorio Astronomico di Cagliari, INAF,via della Scienza 5, I-09047 
Selargius (CA), Italy \\
$^6$ Jodrell Bank Centre for Astrophysics, Alan Turing Building, University of
Manchesters, M13 9PL, UK\\
$^7$ Max-Planck-Institute f\"ur Radioastronomie, Auf Dem H\"ugel 69,
D-53121, Bonn, Germany \\
$^8$ Department of Physics \& Astronomy, University of British Columbia,
6224 Agricultural Road, Vancouver, B.C. V6T 1Z1, Canada \\
$^9$ University of Cambridge, Cavendish Laboratory, JJ Thomson Avenue,
Cambridge, CB3~0HE \\
$^{10}$ Universit\`a degli Studi di Cagliari, Dipartimento di Fisica, SP
   Monserrato-Sestu km 0,7, 90042, Monserrato (CA), Italy \\
$^{11}$ Columbia Astrophysics Laboratory, Columbia University, 550 West 
120th Street, New York, NY 10027, USA \\
$^{12}$ Department of Physics and Astronomy, Franklin \& Marshall College,
PO~Box~3003, Lancaster, PA 17604, USA}
\date{Accepted 2015 April 9. Received 2015 March 6; in original form 2015 January 20}
\let\leqslant=\leq 
\begin{document}

\maketitle
\newcommand{\setthebls}{
}

\setthebls

\begin{abstract} 
We present timing observations of four millisecond pulsars discovered
in the Parkes 20-cm multibeam pulsar survey of the Galactic
plane. PSRs J1552$-$4937 and J1843$-$1448 are isolated objects with
spin periods of 6.28 and 5.47~ms respectively. PSR~J1727$-$2946 is in
a 40-day binary orbit and has a spin period of 27~ms. The 4.43-ms
pulsar J1813$-$2621 is in a circular 8.16-day binary orbit around a
low-mass companion star with a minimum companion mass of
0.2~$M_{\odot}$.  Combining these results with detections from five
other Parkes multibeam surveys, gives a well-defined sample of 56
pulsars with spin periods below 20~ms.  We develop a likelihood
analysis to constrain the functional form which best describes the
underlying distribution of spin periods for millisecond pulsars.  The
best results were obtained with a log-normal distribution. A gamma
distribution is less favoured, but still compatible with the
observations.  Uniform, power-law and Gaussian distributions are found
to be inconsistent with the data. Galactic millisecond pulsars being
found by current surveys appear to be in agreement with a log-normal
distribution which allows for the existence of pulsars with periods
below 1.5~ms.
\end{abstract}


\begin{figure*} 
\centerline{\psfig{file=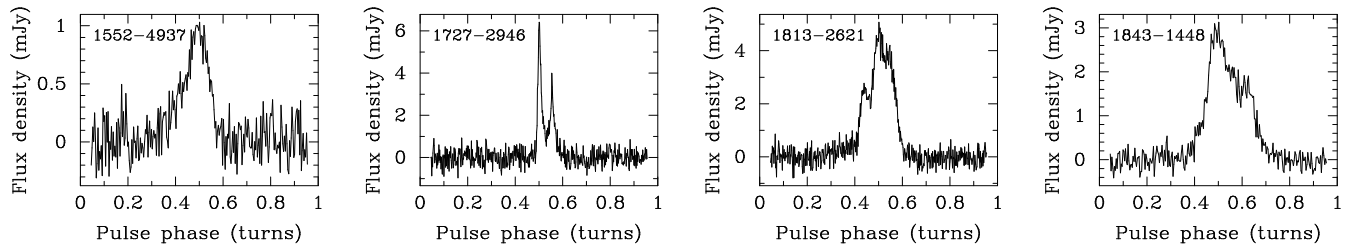,width=\textwidth,angle=270}}
\caption{Integrated pulse profiles each showing 360 degrees of
  rotational phase at 20~cm wavelength for the four MSPs described in
  this paper. Data were obtained with the Parkes digital filterbank
  systems.}
\label{fig:profiles}
\end{figure*}

\begin{figure*} 
\centerline{\psfig{file=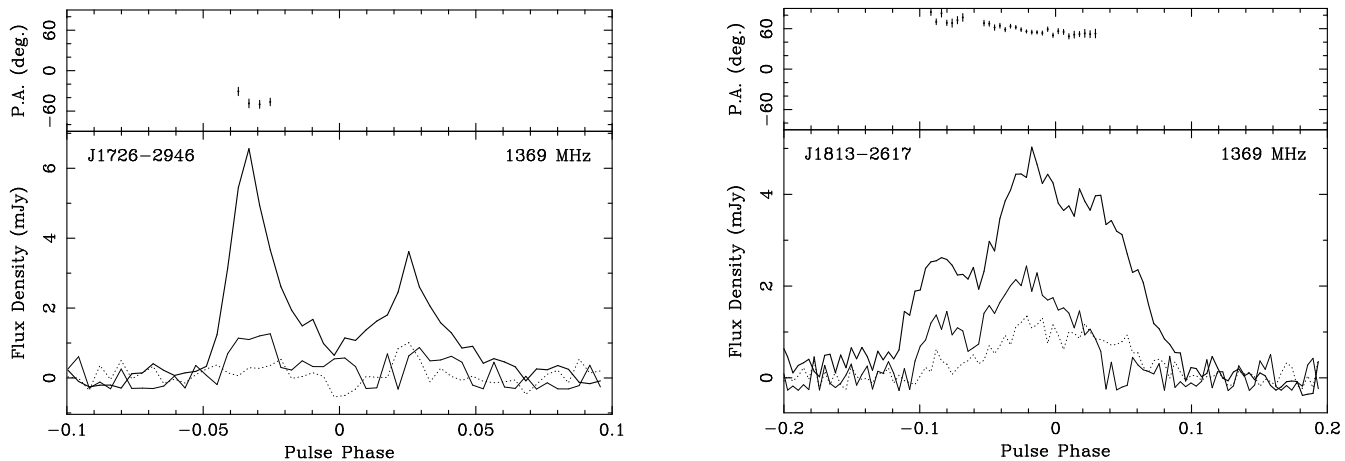,width=\textwidth}}
\caption{Integrated polarization profiles for the two MSPs for which
  significant RMs were detected. The top panels of each plot show the
  polarization position angle as a function of pulse phase. The bottom
  panels show total intensity (bold line), linear polarization (solid
  line) and circular polarization (dotted line).}
\label{fig:pol}
\end{figure*}

\section{INTRODUCTION}

Millisecond radio pulsars (MSPs) are fascinating objects to
study. Their phenomenal rotational stability allows them to be used
for a wide variety of fundamental physics experiments including as a
Galactic-scale observatory to search for low-frequency gravitational
waves \citep[see, e.g.,][]{haa+10}.  Ever since the discovery of the
first MSP \citep{bkh+82} it has been clear that the difficulties in
detection imply that the Galactic population of MSPs is
substantial. Early studies showed that the population of MSPs is
comparable to that of normal pulsars \citep[see,.e.g.,][]{kn88,jb91}.

The continued improvement of data acquisition systems over the past
twenty years has led to a dramatic increase in survey sensitivity to
MSPs. The number of known MSPs in the Galactic disk (i.e.,~those not
associated with globular clusters) is now
230\footnote{\url{http://astro.phys.wvu.edu/GalacticMSPs}}\footnote{\url{http://www.atnf.csiro.au/research/pulsar/psrcat}}.
Because of the great success of blind surveys of the Galactic field
\citep[for a review, see][]{sll13}, and targeted searches of {\it
  Fermi} sources \citep{rap+12}, for the first time in a decade,
Galactic MSPs outnumber their counterparts in globular clusters.

The Parkes multibeam pulsar survey (PMPS) of the Galactic plane is the
most successful large-scale search for pulsars so far undertaken.  Six
previous papers in this series have presented timing parameters for
742 newly discovered pulsars and have discussed various aspects of the
survey results
\citep{mlc+01,mhl+02,kbm+03,hfs+04,fsk+04,lfl+06,cls+13}.  Over the
past five years, several re-analyses of the survey data have been
carried out. \citet{kel+09} discovered a further 28 pulsars by
applying new candidate sorting algorithms to the data processed
earlier by \citet{fsk+04}. \citet{ekl09} applied a new interference
removal technique to a small portion of the data and discovered a
further four pulsars. In a further reanalysis, Keane et al.~found one
fast radio burst \citep{kskl12} and 10 rotating radio transients
\citep{kle+10} in addition to the 11 found originally by
\citet{mll+06}.  \citet{mlb+12} reported the discovery of the 34.5~ms
binary pulsar J1725$-$3853 as well as four other millisecond
pulsars. One other binary MSP, J1753$-$2814, has also been discovered
as a result of this processing effort (Mickaliger et al.~in
preparation). Following earlier discoveries using ``stack-slide''
acceleration searches by \citet{fsk+04}, \citet{eklk13} report the
discovery of 16 pulsars in a coherent acceleration search of the data.
Ongoing processing by Einstein@Home volunteers \citep{kek+13} has
resulted in the discovery of a further 23 pulsars.

In this paper, we present timing solutions for four MSPs discovered in
the PMPS. Preliminary discovery and confirmation observations of these
pulsars were previously published by \citet{fsk+04}. The total number
of pulsars found in the survey so far stands at 833. Since extensive
population studies of the normal pulsar population as revealed by the
PMPS have already been carried out \citep{fk06,lfl+06}, in this paper
we focus our discussion on the spin period distribution of the MSP
population.  The plan for this paper is as follows.  In
\S\ref{sec:timing} we present the basic timing parameters, pulse
widths, mean profiles and flux densities for the four new MSPs. In
\S\ref{sec:psrpop} we compile a sample of MSPs and use it to carry out
a likelihood analysis to constrain the underlying distribution of spin
periods. The main conclusions are summarized in \S\ref{sec:conc}.

\nocite{cl02}
\begin{table*}
\label{tab:timing}
\centering
\caption{Spin, astrometric and derived parameters from the timing
  analysis of four MSPs. All astrometric parameters are given in the
  J2000 coordinate system. The reduced $\chi^2$ values from each fit
  is listed as $\chi^2_r$. Figures in parentheses represent the
  uncertainties in the least significant digit and are the nominal
  1-$\sigma$ {\sc tempo2} uncertainties. Distance estimates are based
  on the pulsar DM using the Cordes \& Lazio (2002) NE2001
  electron-density model. Pseudo-luminosities are computed by
  multiplying flux density by distance squared. Characteristic ages,
  magnetic fields and spin-down luminosities are based on the spin
  period and period derivative (see, e.g., Lorimer \& Kramer 2005) and
  account for the contributions due to the Shklovskii effect and
  Galactic acceleration.}
\begin{tabular}{@{}lllll}
\hline
Parameter & PSR~J1552$-$4937 & PSR~J1727$-$2946 & PSR~J1813$-$2621& PSR~J1843$-$1448\\
\hline
R.A.  (hh:mm:ss.s) & 15:52:13.2709(4)&17:27:15.09493(17) &18:13:40.59165(10)&
18:43:01.3750(3)\\
Dec. (dd:mm:ss.s) & $-$49:37:49.744(11)&$-$29:46:36.797(17)&$-$26:21:57.055(18)&
$-$14:48:12.61(3)\\
Proper motion in R.A.  (mas~y$^{-1}$) & $-$3(3)& 0.6(9)& $-$7.3(9) & 10.5(19)\\
Proper motion in Dec.  (mas~y$^{-1}$) & $-$13(8)& 0(8)& $-$22(16) & 12(15)\\
Epoch of position                     & J2000   & J2000 & J2000 & J2000 \\
Spin period (ms) & 6.2843113814174(12)&27.0831832440066(12) &4.4300116286341(18)&
5.4713308755095(6)\\
First derivative of spin period 
& $1.900(4)\times10^{-20}$&$2.4632(3)\times10^{-19}$
&$1.2466(6)\times10^{-20}$&$6.209(18)\times10^{-21}$\\
Dispersion measure (cm$^{-3}$ pc) & 114.19(8)&60.74(3)&112.524(9)&114.51(7)\\
Rotation measure (rad~m$^{-2}$) &  --- & --61(32) & 136(8) & ---\\
Epoch of period (MJD) & 54033& 54723 &54058&53934\\
Data span (MJD) & 52860--55206 & 52666--56781 & 52696--55419 & 52696--55152\\
$\chi^2_r$ / degrees of freedom &  1.00/132& 1.45/208 &0.90/134&1.16/115\\
Post-fit rms residual ($\mu$s) & 78& 43 &17.5&49\\
\hline
Flux density at 1.4 GHz (mJy) & 0.14 & 0.25 & 0.65 & 0.57\\
Pulse width at 50\% of peak (ms) & 0.9 & 1.8 & 0.66 & 1.0 \\
Distance (kpc) & 4.8 & 1.4 & 2.9 & 2.9\\
Pseudo-luminosity (mJy~kpc$^2$) & 3.2 & 0.49 & 2.1 & 4.8\\
Intrinsic period derivative & $0.7(1.1)\times10^{-21}$&$2.426(7)\times10^{-19}$&
$-0.6(1.7)\times10^{-21}$&$-0.5(1.1)\times10^{-21}$\\
Characteristic age (Gy) & $>15$  & 1.77 &$>5.6$&$>14$\\
Surface magnetic field ($10^8$~G) & $<2.1$ & 25.9 & $<2.4$ & $<1.9$\\
Spin-down luminosity ($10^{33}$~ergs~s$^{-1}$)&$<1.1$&0.48 & $<5.7$ & $<1.5$ \\
\hline
\end{tabular}
\end{table*}

\section{FOUR MILLISECOND PULSARS}\label{sec:timing}

The pulsars were discovered using the processing schemes described by
\citet{fsk+04}.  Following the confirmation and positional refinement
procedures described by \citet{mhl+02}, each pulsar was observed
regularly at Parkes using initially the $512 \times 0.5$~MHz analogue
filterbank system \citep{mlc+01} and subsequently the digital
filterbank systems \citep{mhb+13}.  For each pulsar, pulse times of
arrival were determined from the individual observations using
standard pulsar timing techniques \citep[see, e.g.,][]{lk05}
implemented in the \textsc{psrchive} software package
\citep{hvm04}\footnote{\url{http://psrchive.sourceforge.net}}. A model
containing the spin, astrometric and (if necessary) any binary
parameters was fitted to the arrival times using the {\sc tempo2}
timing package \citep{hem06}. Arrival times were referred to TT(TAI)
and the DE421 planetary ephemeris \citep{fwb08} was used. Timing
parameters are expressed in ``TCB'' units native to {\sc tempo2}
\citep[see][for the definition of TCB]{hem06}.  Integrated pulse
profiles are shown in Fig.~\ref{fig:profiles}.

Timing parameters from these analyses along with various derived
quantities are presented in Tables~\ref{tab:timing} and
\ref{tab:binary}. For PSR~J1727--2947, time-of-arrival uncertainties
were multiplied by a factor ranging between 0.85--1.5 for different
backend systems to maintain a reduced $\chi^2$ value close to unity.
Also listed in Table~\ref{tab:timing} is the post-fit root-mean-square
residual.  The values obtained from our timing so far are relatively
large (17--78~$\mu$s) and indicate that these pulsars are unlikely to
be useful additions to MSP timing arrays. Although proper motions in
right ascension have been measured for PSRs J1813$-$2621 and
J1843$-$1448, we are unable to measure a significant proper motion in
declination because of the low ecliptic latitude of these
pulsars. Flux densities at 1400~MHz and pulse widths at 50\% of the
peak level based on the profiles shown in Fig.~\ref{fig:profiles} are
listed in Table~\ref{tab:timing}.
 
For two of the MSPs, J1727--2947 and J1813--2621, significant levels
of polarized emission was measured and these are shown in the
integrated pulsed profiles in Fig.~\ref{fig:pol}. Rotation measures
for both these pulsars were determined using the {\tt rmfit} tool
within \textsc{psrchive} with conservative estimates of the
uncertainties.

PSRs~J1552$-$4937 and J1843$-$1448 bring the total number of isolated
MSPs known in the Galactic disk to 37.  When compared to the sample of
172 MSPs for which an orbiting companion has been confirmed, the
fraction of observed isolated MSPs currently stands at 18\%.  An
outstanding issue in our understanding of MSP population is to explain
this population in a self-consistent fashion. In particular, an open
question is whether isolated MSPs formed in a different way from
binary MSPs.  We discuss this issue further in \S~\ref{sec:discuss}.

PSR~J1727$-$2947 is a relatively long-period MSP ($P \sim 27$~ms) in a
mildly eccentric ($e \sim 0.04$) 40-day binary system. With a minimum
companion mass of $\sim 0.8$~$M_\odot$, the system is most likely a
member of the so-called ``intermediate-mass binary pulsar'' (IMBP)
class \citep{clm+01} with a relatively massive CO white dwarf
companion.  The parameters for PSR~J1813$-$2621 imply that it is very
representative of the low-mass binary MSP population. Interpreting the
orbital parameters in the standard way \cite[see, e.g.,][]{lk05}, we
infer a companion mass of at least 0.2~$M_{\odot}$, typical of a
low-mass white dwarf. Optical studies of these companions may provide
further insights into the nature of these two binary systems.

For nearby MSPs, it is well known \citep[see, e.g.,][]{dt91} that two
significant contributions to the observed period derivative are the
effects of secular acceleration \cite[sometimes referred to as the
  ``Shklovskii effect'',][]{shk70} and Galactic acceleration. For a
pulsar of period $P$, transverse speed $V$, acceleration ${\bf a}$,
distance $D$ with an intrinsic period derivative $\dot P_{\rm int}$,
the observed period derivative
\begin{equation}
\dot{P}_{\rm obs} = \dot{P}_{\rm int} + P \left(
\frac{{\bf a}\cdot{\bf {\hat n}}}{c} + 
\frac{V^2}{cD}
\right),
\end{equation}
where ${\bf {\hat n}}$ is a unit vector along the line of sight to the
pulsar and $c$ is the speed of light. Following the discussion in \S
3.1 of \citet{nt95} to compute these effects, and assuming a 25\%
uncertainty on the distances, computed using the NE2001 electron
density model \cite{cl02}, we calculated or placed limits on
$\dot{P}_{\rm int}$ for each pulsar and list our results in Table~1.
The resulting characteristic age and magnetic field strength estimates
for these pulsars are also indicated in Table~1. As can be seen, for
all but PSR~J1727--2946, these effects account for most of the
observed period derivative.

\nocite{lcw+01}
\begin{table}
\label{tab:binary}
\centering
\caption{Measured and derived orbital parameters for PSRs~J1727$-$2946
  and J1813$-$2621 which use the ``BT'' binary model \citep{bt76} and
  the ``ELL1'' binary model (Lange et al.~2001)
  respectively. Parameters listed are the binary period ($P_b$),
  projected semi-major axis ($a \sin i$), orbital eccentricity ($e$),
  first and second Laplace-Lagrange parameters ($\epsilon_1$ and
  $\epsilon_2$), longitude and epoch of periastron ($\omega$ and
  $T_0$) and epoch of ascending node ($T_{\rm asc}$).  Figures in
  parentheses represent 1-$\sigma$ {\sc tempo2} uncertainty in the
  least significant digits. For PSR J1813$-$2621, we also list the
  corresponding values of $e$, $\omega$ and $T_0$ computed from the
  Laplace-Lagrange parameters.  The Keplerian mass function ($4\pi^2 G
  (a \sin i)^3/P_b^2$, where $G$ is Newton's constant), and the
  minimum companion mass (calculated assuming a 1.4~$M_{\odot}$ pulsar
  and setting, $i = 90\degr$) are listed.}
\begin{tabular}{@{}lll}
\hline
PSR          & J1727$-$2946 & J1813$-$2621 \\
\hline
Binary model & BT & ELL1 \\
$P_b$ (d)     & 40.30771094(3) & 8.159760702(10)\\
$a1\sin i$ (lt sec) & 56.532497(5)  & 5.592583(3)\\
$e$       &0.04562943(16) & $2.7^{+1.2}_{-1.1}\times 10^{-6}$ \\
$\omega$ ($\degr$) & 320.39625(20) & $289_{-18}^{+28}$ \\
$T_0$ (MJD)  & 54711.47169(3) & $54061.5_{-4.0}^{+6.0}$ \\
$\epsilon_1$    & -- &$-2.5(10)\times 10^{-6}$\\
$\epsilon_2$    & -- & $9(8)\times 10^{-7}$\\
$T_{\rm asc}$ (MJD)  & -- & 54054.9328319(6)\\
\hline
Mass function ($M_{\odot}$)       &   0.1194  &   0.00282       \\
Min. comp. mass ($M_{\odot}$) &   0.827   &  0.188 \\ 
\hline
\end{tabular}
\end{table}

\section{The spin period distribution of Galactic MSPs}\label{sec:psrpop}

\begin{table} 
\label{tab:sample}
\centering
\caption{The 56 MSPs used in the population study. For each pulsar, we
  list the spin period ($P$), dispersion measure (DM), Galactic
  longitude ($l$), Galactic latitude ($b$), whether this is a binary
  pulsar as well as the survey which detected the pulsar. The surveys
  considered were the PMPS, the Parkes high-latitude (PH) pulsar
  survey, the Perseus arm (PA) pulsar survey, the Deep Multibeam (DMB)
  survey, the Swinburne intermediate latitude (SWIL) and high latitude
  (SWHL) surveys.}
\begin{tabular}{@{}lrrrrrr}
\hline
PSRJ & \multicolumn{1}{c}{$P$} & \multicolumn{1}{c}{DM} & \multicolumn{1}{c}{$l$} & \multicolumn{1}{c}{$b$} & Bin & Survey \\
    & \multicolumn{1}{c}{(ms)}& \multicolumn{1}{c}{(pc/cc)} & \multicolumn{1}{c}{($^{\circ}$)}  & \multicolumn{1}{c}{($^{\circ}$)} &   & \\
\hline 
0437$-$4715 & 5.76 & 2.6 & 253.4 & $-42.0$ & Y& PH\\
0610$-$2100 & 3.86 & 60.7 & 227.7 & $-18.2$ &Y & PH\\
0711$-$6830 & 5.49 & 18.4 & 279.5 & $-23.3$ &N & SWHL\\
0721$-$2038 & 15.54 & 76.1 & 234.7 & $-2.9$ &Y & PA\\
0900$-$3144 & 11.11 & 75.7 & 256.2 & $9.5$ & Y& PH\\
0922$-$52 & 9.68 & 122.4 & 273.8 & $-1.4$ & Y& PMPS\\
1022$+$1001 & 16.45 & 10.2 & 231.8 & $51.1$ &Y & PH\\
1024$-$0719 & 5.16 & 6.5 & 251.7 & $40.5$ & N& PH\\
1045$-$4509 & 7.47 & 58.2 & 280.9 & $12.3$ & Y& SWIL\\
1125$-$6014 & 2.63 & 53.0 & 292.5 & $0.9$ & Y& PMPS\\
1147$-$66 & 3.72 & 133.5 & 296.5 & $-4.0$ & Y& PMPS\\
1216$-$6410 & 3.54 & 47.4 & 299.1 & $-1.6$ & Y& PMPS\\
1435$-$6100 & 9.35 & 113.7 & 315.2 & $-0.6$ &Y & PMPS\\
1546$-$59 & 7.79 & 168.2 & 323.5 & $-3.8$ & Y& PMPS\\
1552$-$4937 & 6.28 & 114.6 & 330.0 & $3.5$ & N& PMPS\\
1600$-$3053 & 3.60 & 52.3 & 344.1 & $16.5$ & Y& SWIL\\
1603$-$7202 & 14.84 & 38.0 & 316.6 & $-14.5$ &Y & SWIL\\
1618$-$39 & 11.99 & 117.5 & 340.8 & $7.9$ & Y& SWIL\\
1629$-$6902 & 6.00 & 29.5 & 320.4 & $-13.9$ & N& SWIL\\
1643$-$1224 & 4.62 & 62.4 & 5.7 & $21.2$ & Y& SWHL\\
1652$-$48 & 3.78 & 187.8 & 337.9 & $-2.9$ & Y& PMPS\\
1708$-$3506 & 4.50 & 146.8 & 350.5 & $3.1$ & Y& PMPS\\
1713$+$0747 & 4.57 & 16.0 & 28.8 & $25.2$ & Y& SWHL\\
1721$-$2457 & 3.50 & 47.8 & 0.4 & $6.8$ & N& SWIL\\
1723$-$2837 & 1.86 & 19.9 & 357.3 & $4.2$ & Y& PMPS\\
1725$-$38 & 4.79 & 158.4 & 349.4 & $-1.8$ & Y& PMPS\\
1730$-$2304 & 8.12 & 9.6 & 3.1 & $6.0$ & N& SWIL\\
1732$-$5049 & 5.31 & 56.8 & 340.0 & $-9.5$ & Y& SWIL\\
1738$+$0333 & 5.85 & 33.8 & 27.7 & $17.7$ & Y& SWHL\\
1741$+$1351 & 3.75 & 24.0 & 37.9 & $21.6$ & Y& SWHL\\
1744$-$1134 & 4.07 & 3.1 & 14.8 & $9.2$ & N& SWIL\\
1745$-$0952 & 19.38 & 64.5 & 16.4 & $9.9$ & Y& SWIL\\
1748$-$30 & 9.68 & 420.2 & 359.2 & $-1.1$ & Y& PMPS\\
1751$-$2857 & 3.92 & 42.8 & 0.6 & $-1.1$ & Y& PMPS\\
1753$-$2814 & 18.62 & 298.4 & 1.4 & $-1.2$ & Y& PMPS\\
1757$-$5322 & 8.87 & 30.8 & 339.6 & $-14.0$ & Y& SWIL\\
1801$-$1417 & 3.62 & 57.2 & 14.5 & $4.2$ & N& PMPS\\
1801$-$3210 & 7.45 & 176.7 & 358.9 & $-4.6$ &Y & PMPS\\
1802$-$2124 & 12.65 & 149.6 & 8.4 & $0.6$ & Y& PMPS\\
1804$-$2717 & 9.34 & 24.7 & 3.5 & $-2.7$ & Y& PMPS\\
1813$-$2621 & 4.43 & 122.5 & 5.3 & $-3.9$ & Y& PMPS\\
1826$-$24 & 4.70 & 81.9 & 8.3 & $-5.7$ & Y& PMPS\\
1835$-$0115 & 5.12 & 98 & 29.9 & $3.0$ & Y& PMPS\\
1843$-$1113 & 1.85 & 60.0 & 22.1 & $-3.4$ &N & PMPS\\
1843$-$1448 & 5.47 & 114.6 & 18.9 & $-4.8$ &N & PMPS\\
1853$+$1303 & 4.09 & 30.6 & 44.9 & $5.4$ & Y& PMPS\\
1857$+$0943 & 5.36 & 13.3 & 42.3 & $3.1$ & Y& PMPS\\
1905$+$0400 & 3.78 & 25.7 & 38.1 & $-1.3$ & N& PMPS\\
1909$-$3744 & 2.95 & 10.4 & 359.7 & $-19.6$ &Y & SWHL\\
1910$+$1256 & 4.98 & 38.1 & 46.6 & $1.8$ &Y & PMPS\\
1911$+$1347 & 4.63 & 31.0 & 47.5 & $1.8$ & N& PMPS\\
1918$-$0642 & 7.65 & 26.6 & 30.0 & $-9.1$ & Y& SWIL\\
1933$-$6211 & 3.54 & 11.5 & 334.4 & $-28.6$ &Y & SWHL\\
1934$+$1726 & 4.20 & 62.0 & 53.2 & $-1.1$ & Y& DMB\\
1939$+$2134 & 1.56 & 71.0 & 57.5 & $-0.3$ & N& DMB\\
2010$-$1323 & 5.22 & 22.2 & 29.4 & $-23.5$ & Y& SWHL\\
\hline 
\end{tabular}
\end{table}

The large sample of over 1000 normal pulsars detected in the various
Parkes multibeam surveys has provided significant advances in our
understanding of the normal pulsar population \citep[see,
  e.g.,][]{fk06,lfl+06}. The observed sample of MSPs is substantially
less numerous because of their generally lower luminosity and
observational selection effects. Nevertheless, the large sky coverage
and uniformity of the observing systems of the multibeam surveys
provides an excellent sample to begin characterizing their
population. Recent work by \citet{lor13} models this population via
Monte Carlo realizations of synthetic pulsars drawn from distribution
functions. These synthetic populations are subsequently ``observed''
with realistic models of the surveys to produce samples that can be
compared with the observed data. In this paper, we focus on
constraining the spin period distribution of MSPs.

For this study, we define a MSP as a pulsar with $P<20$~ms. Our final
sample of 56 MSPs is drawn from detections by this survey (PMPS), the
Swinburne Intermediate Latitude Survey \citep[SWIL;][]{ebvb01}, the
Swinburne High Latitude Survey \citep[SWHL;][]{jbo+07}, the Parkes
High Latitude Survey \citep[PH;][]{bjd+06}, the Perseus Arm Survey
\citep[PA;][]{bkl+13}, and the Deep Multibeam Survey
\citep[DMB;][]{lcm13}. The basic parameters of this sample of pulsars
are summarized in Table~\ref{tab:sample}. For the purposes of
population analyses, this sample of pulsars is a very natural one to
analyse, since the pulsars were found using the same telescope,
receiver and data acquisition system. Since the sensitivity of this
system is well understood \citep{mlc+01,lfl+06}, our survey models are
reliable.  In addition, since the surveys were all carried out at
20~cm wavelength, we need not make assumptions about MSP flux density
spectra in order to extrapolate results from surveys carried out at
other frequencies.

\subsection{Likelihood analysis description}\label{sec:likedescribe}

Following earlier work \citep[e.g.,][]{cc97}, we adopt a likelihood
analysis to constrain the period distribution. In our approach to this
problem, the probability $p_i$ of detecting pulsar $i$ in the sample
with period $P_i$, and dispersion measure DM$_i$ can be written as
follows:
\begin{equation}
\label{equ:detpdf}
  p_i = f(P|\,a,b) {\cal D}(P_i,{\rm DM}_i).
\end{equation}
In this expression, as usual in statistical parlance, the ``$|$''
symbol denotes a conditional probability. The quantity $f(P|\,a,b)$
represents the probability density function (PDF) of the period which
we seek to constrain.  All the models considered in this work can be
described by two parameters which we refer to here generally as $a$
and $b$. Specific parameters will be defined below. The detectability
function, ${\cal D}$, reflects the probability of detecting the pulsar
in one of the six surveys mentioned above and therefore appropriately
accounts for their non-uniform period sensitivity. Note that in this
analysis, we assume that any loss of sensitivity due to binary motion
is negligible given the significant number of acceleration searches
that have been carried out on these data \citep{fsk+04,kek+13,eklk13}.

Given a model period PDF and a detectability model which we describe
in detail below, the likelihood function ${\cal L}(a,b)$ for a given
combination of $a$ and $b$ is simply the product of all the 56
individual $p_i$ values.  The optimal set of model parameters
$\hat{a}$ and $\hat{b}$ are those which maximize ${\cal L}$ in a grid
of parameter space over $a$ and $b$.  Once the maximum likelihood
${\cal L}_{\rm max}$ has been found, the marginalized PDF for $a$ is
then simply the distribution of ${\cal L}(a,\hat{b})$ (and vice versa
for the PDF of $b$). This approach provides PDFs for $a$ and $b$ as
well as a means for evaluating different period PDFs from the ratio of
the maximum likelihoods (i.e., the Bayes factor, $K$). For example,
given two model PDFs ``$x$'' and ``$y$'', $K={\cal L}_{\rm
  max,x}/{\cal L}_{\rm max,y}>1$ if model $x$ better describes the
sample than model $y$.  According to Jeffreys (1961), \nocite{jef61} a
Bayes factor of between 1 and 3 is deemed to be essentially
indistinguishable from the best model while Bayes factors in the range
3--10 begin to favour $x$ over $y$.  Bayes factors higher than 100
decisively favour $x$.

\subsection{Detectability model}\label{sec:detectability}

The detectability of a given pulsar, ${\cal D}$, reflects how likely
it is to be found in the sample of 56 MSPs we will ultimately be
applying this analysis to. Calculating ${\cal D}$ therefore requires
accurately accounting for the difficulties in detecting each
pulsar. Two approaches that can be brought to bear on this problem are
to: (i) run large numbers of Monte Carlo simulations which model the
detectability; (ii) develop a simple analytical model. After initial
experimentation with the first approach, it became clear to us that
the Monte Carlo simulations require a large number of assumptions and
significant computational resources to carry out a sufficient number
of realizations necessary to estimate ${\cal D}$. We therefore
followed the second approach and calculate ${\cal D}$ for each MSP in
our sample.  The essence of our approach, described in detail below,
is to find for any line of sight the flux density distribution of
pulsars: $p(S)$. The detectability of a pulsar along this line of
sight is then the fraction of such pulsars visible by a survey.  For
the $i^{\rm th}$ pulsar, we may therefore write
\begin{equation}
\label{eq:detectability}
  {\cal D}_i = \frac{\int_{S_{{\rm min},i}}^{\infty} p(S) {\rm d}S}
                  {\int_{S_0}^{\infty} p(S) {\rm d}S}.
\end{equation}
Here $S_{{\rm min},i}$ is the minimum detectable survey flux density
of this pulsar which we compute from its pulse period, dispersion
measure and pulse width (an intrinsic pulse duty cycle of 10\% was
assumed for each pulsar). The term $S_0$ represents the lowest flux
density detectable in the survey, i.e the limit of $S_{{\rm min},i}$
as $P$ becomes large and DM tends to zero.  In paper VI of this series
(Lorimer et al.~2006), we gave expressions for computing $S_{{\rm
    min},i}$ and refer the reader to this work for details. The
advantage of this approach to the problem is that it does not depend
on knowledge of individual distances to MSPs, which are uncertain. The
analysis also does not depend on detected signal-to-noise or flux
densities for any individual pulsars.  Instead it takes advantage of
prior information about the pulsar population to calculate $p(S)$
rigorously. As we show below, the final determination of the
detectability function for this sample of pulsars can be given in
terms of only two parameters which are robust to uncertainties
inherent in the assumptions.

\begin{figure} 
\centerline{\psfig{file=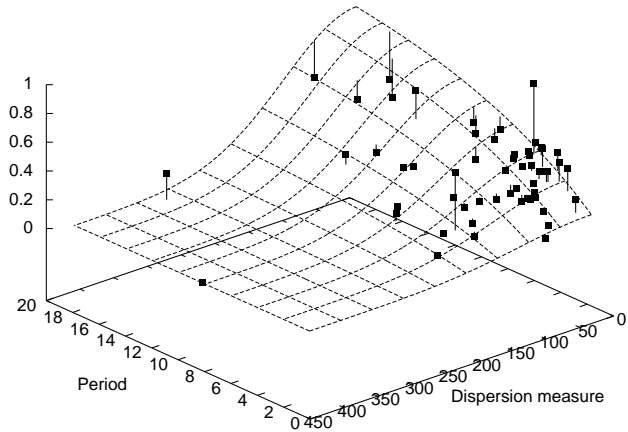,width=10cm,angle=270}} 
\caption{Detectability as a function of period (in ms) and dispersion
  measure (in cm$^{-3}$~pc). The data points show our estimates of
  detectability for each pulsar which we compute as described in the
  text. The grid shows our best approximation to this behaviour using
  the two-dimensional detectability function defined in equation
  \ref{eq:det} assuming the parameters $\alpha=10$~ms and
  $\beta=110$~cm$^{-3}$~pc.  For each data point, a vertical line is
  drawn showing its distance from the best-fitting surface.}
\label{fig:detectability}
\end{figure}

The calculation of $p(S)$ is most readily achieved from an application
of Bayes' theorem which implies for a given line of sight the
following relationship between PDFs in flux density $S$ and distance $D$:
\begin{equation}
\label{equ:bayes}
  p(D|S) \propto p(S|D) p(D).
\end{equation}
Because the distribution we seek, $p(S)$, is simply $p(S|D)$
marginalized over distance, we can use Eq.~\ref{equ:bayes} to show that
\begin{equation}
\label{eq:ps}
p(S) = \int_0^{\infty} p(S|D) {\rm d}D \propto \int_0^\infty 
\frac{p(D|S)}{p(D)} {\rm d}D.
\end{equation}
To get expressions for $p(D|S)$ and $p(D)$, we use the results
described in \S 3.3 of \citet{vwc+12}. In this work, assuming a
log-normal pulsar luminosity function \citep[see, e.g.,][]{fk06} with
mean $\mu$ and standard deviation $\sigma$, it is shown that
\begin{equation}
  p(D|S) \propto \frac{1}{D} \exp \left[ -\frac{1}{2} \left(
  \frac{\log S + 2 \log D - \mu}{\sigma} \right)^2 \right].
\end{equation}
\citet{vwc+12} also show that, along a line of sight defined by
Galactic longitude $l$ and latitude $b$, an axisymmetric distribution
of pulsars with the radial density profile found in paper VI leads to
the result
\begin{equation}
p(D)\propto R^{1.9} D^2 \exp\left[-\frac{|z|}{h}\right]
                        \exp\left[-\rho \frac{|R-R_0|}{R_0}\right].
\end{equation}
In this expression, $z = D \sin b$ is the vertical height off the
Galactic plane, $h$ is the scale height of pulsars, $\rho$ is a
dimensionless parameter used to scale the population over the Galactic
disk, $R_0=8.5$~kpc is the Galactocentric radius of the Sun and the
pulsar Galactocentric radius
\begin{equation}
R = \sqrt{R_0^2 + (D \cos b)^2 - 2 R_0 D \cos b \cos l}.
\end{equation}
With these analytical results it is then straightforward using
numerical integration of Eq.~\ref{eq:ps} to find the appropriate form
of $p(S)$ for each $l$ and $b$. This PDF is then numerically
integrated according to the limits in Eq.~\ref{eq:detectability} to
find ${\cal D}_i$.

Following the results of \citet{blc11} and Lorimer (2013), we adopted
nominal parameter values of $h=500$~pc, $\rho=5$, $\mu=-1.1$ and
$\sigma=0.9$.  Fig.~\ref{fig:detectability} shows the results of this
calculation on our sample of 56 pulsars as scatter plots of ${\cal D}$
as a function of $P$ and DM. As expected, ${\cal D}$ is lower for
shorter period and/or higher DM pulsars. To approximate this trend in
the likelihood analysis, we set
\begin{equation}
\label{eq:det}
  {\cal D} = [1 - \exp(-P/\alpha)] \, \exp(-{\rm DM}^2/2 \beta^2),
\end{equation}
where simple fits to the data show that $\alpha \simeq 10$~ms and
$\beta \simeq 110$~cm$^{-3}$~pc provide a good description of these
trends, as shown by the smooth surface in
Fig.~\ref{fig:detectability}.  The root-mean-square deviation of the
data from this surface is 0.1.  As described in \S \ref{sec:validity},
while the choice of duty cycle or population parameters assumed in the
detectability analysis impacts the values of $\alpha$ and $\beta$
somewhat, the particular values of $\alpha$ and $\beta$ do not
significantly affect our conclusions.

\begin{figure*} 
\centerline{\psfig{file=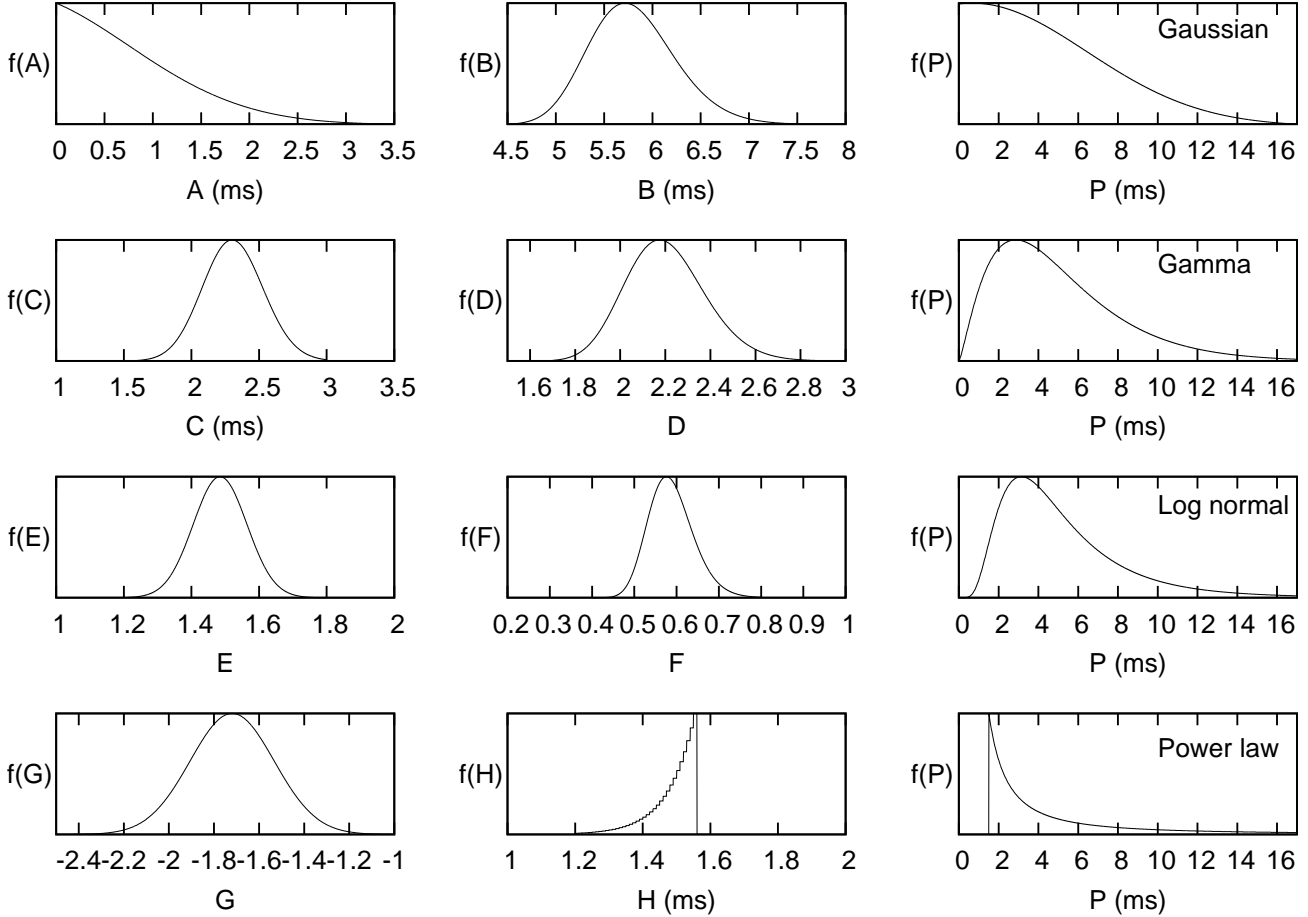,width=18cm,angle=270}} 
\caption{Left and centre columns: marginalized posterior PDFs for each
  of the model parameters A---H (see equations
  \ref{eq:gauss}---\ref{eq:power} for definitions) obtained from the
  likelihood analysis described in the text.  Right column:
  corresponding PDF for $f(P)$ for each distribution when the nominal
  parameter values given in Table~4 are adopted.}
\label{fig:marginal}
\end{figure*}

\subsection{Period distributions investigated}

We considered a variety of analytical functions to find the PDF which
best describes the spin period of the MSP population. The simplest
case of a uniform distribution is clearly not favoured by the data. In
preliminary investigations we found Bayes factors relative to other
models of order $10^{-12}$ and disregarded it from further
analysis. Better approximations to the true period PDF can be found by
considering functions with some well defined peak.  All four
functional forms we investigate henceforth (i.e.,~Gaussian, gamma,
log-normal and power law distributions) require two parameters. In a
similar way to paper VI, we refer to these parameters using capital
letters.  The Gaussian distribution has a mean $A$ and standard
deviation $B$:
\begin{equation}
\label{eq:gauss}
  f(P)_{\rm gauss} \propto \exp\left[\frac{-(P-A)^2}{2B^2}\right].
\end{equation}
The gamma distribution is parameterized by $C$ and $D$:
\begin{equation}
\label{eq:gamma}
  f(P)_{\rm gamma} \propto \exp(-P/C) (P/C)^{D-1}.
\end{equation}
The log-normal distribution is parameterized by $E$ and $F$:
\begin{equation}
\label{eq:lnorm}
  f(P)_{\rm lnorm} \propto \frac{1}{P} 
\exp \left[\frac{-(\ln(P)-E)^2}{2F^2}\right].
\end{equation}
We also considered a power-law distribution parameterized by an 
exponent $G$ and a minimum period $H$ as follows:
\begin{eqnarray}
  f(P)_{\rm power} & =       & 0 \,\,\, {\rm for} \,P \leq H,   \\
  f(P)_{\rm power} & \propto & P^G \,\,\, {\rm for} \,P>H \,\, {\rm and} 
  \,P < 20~{\rm ms},\\
  f(P)_{\rm power} & =       & 0 \,\,\, {\rm for} \,P \geq 20~{\rm ms}.
\label{eq:power}
\end{eqnarray}
Note that the last boundary condition simply reflects our definition
of a MSP as a pulsar with $P<20$~ms.

\subsection{Application to the observed sample}

Using the above method, we maximize ${\cal L}$ for each of the period
distribution models. A program was
written\footnote{\url{http://psrpop.phys.wvu.edu/pdist}} to implement
the analysis and derive marginalized PDFs of the resulting model
parameters.  For each model, we normalized the detection probability
and period distribution such that
\begin{equation}
\int_0^{\infty} {\cal D}(P,{\rm DM}) \, f(P|\,a,b) \,\,{\rm d}P = 1.
\end{equation}
This normalization ensures that the resulting likelihood values can be
compared with one another to compute Bayes factors. In the results
below, we give the Bayes factors for the best model relative to each
model under consideration.

\begin{table}
\centering
\caption{Results of the likelihood analysis for the period
  distribution models considered. For each model we list the median
  and 95\% confidence interval on the parameters defined in equations
  \ref{eq:gauss}---\ref{eq:power} along with the Bayes factor ($K$)
  computed by dividing the log-normal model likelihood by the
  likelihood of that model.}
\label{tab:results}
\begin{tabular}{lllr}
\hline
Model     & First parameter          & Second parameter          & $K$\\
\hline
Gaussian  & $A=0.7_{-0.6}^{+1.7}$~ms & $B=5.8_{-0.8}^{+1.0}$~ms  & 738 \\
Gamma     & $C=2.3\pm0.4$~ms         & $E=2.2_{-0.3}^{+0.4}$     &  13 \\
Log-normal& $E=1.5\pm0.2$            & $F=0.58_{-0.09}^{+0.12}$  &   1 \\
Power-law & $G=-1.7\pm0.4$           & $H=1.51_{-0.20}^{+0.05}$~ms& 182\\
\hline
\end{tabular}
\end{table}

The results of our analysis when applied to the observed sample of 56
pulsars are summarized in Fig.~\ref{fig:marginal} and
Table~\ref{tab:results}. Fig.~\ref{fig:marginal} shows the
marginalized posterior PDFs for each of the model parameters. 
Table~\ref{tab:results} lists the 95\% credible intervals for all the
model parameters. The highest likelihood values were obtained for the
log-normal model. The Bayes factors of the other models relative to
this one are also given in Table~\ref{tab:results}.  These results
indicate that the log-normal and gamma distributions give by far the
most plausible descriptions of the MSP spin period distribution.

\subsection{Testing the validity of the analysis}
\label{sec:validity}

Before discussing the impact of our results, it is important to
demonstrate the reliability of the parameter estimation approach and
its sensitivity to assumptions. To do this, we generated fake samples
of detectable pulsars with known period distributions and passed these
as input to the likelihood analysis. We used the {\tt psrpop} software
package\footnote{\url{http://psrpop.sourceforge.net}} introduced in
paper VI \citep[see also][]{lor13} to generate synthetic populations
of MSPs for this purpose.  As a starting point we distributed the
model pulsars with model parameter values of $h=500$~pc, $\rho=5$,
$\mu=-1.1$ and $\sigma=0.9$.  For the period distribution, we then
chose each of the four distributions in turn and set the parameters
A--H to be the notional values given in Table~\ref{tab:results} from
our analysis of the real data.  In each simulation, we generated
enough synthetic pulsars such that a total of 56 of them were
detectable by models of the PMPS, Parkes high-latitude (PH), Perseus
arm (PA), Deep Multibeam (DMB), Swinburne intermediate latitude (SWIL)
and Swinburne high latitude (SWHL) surveys available in {\tt psrpop}.
With each synthetic sample, we first carried out a detectability
analysis as described in \S \ref{sec:detectability} to determine
values of the detectability-model parameters $\alpha$ and $\beta$ and
then applied these in our likelihood analysis.  We found that the
returned parameter values A--H from the likelihood analysis were
entirely consistent with the input values of the period distribution
of the parent population. In addition, we found that the method
consistently favored the correct form of the input distribution by
assigning it the maximum likelihood. For example, when we generated
synthetic populations assuming a log-normal distribution, we
consistently found the Bayes factors for the log-normal likelihood
model to be lower than the other distributions, as is seen for the
actual sample of MSPs. Similar results were found when other
underlying period distributions were assumed.

\begin{figure*} 
\centerline{\psfig{file=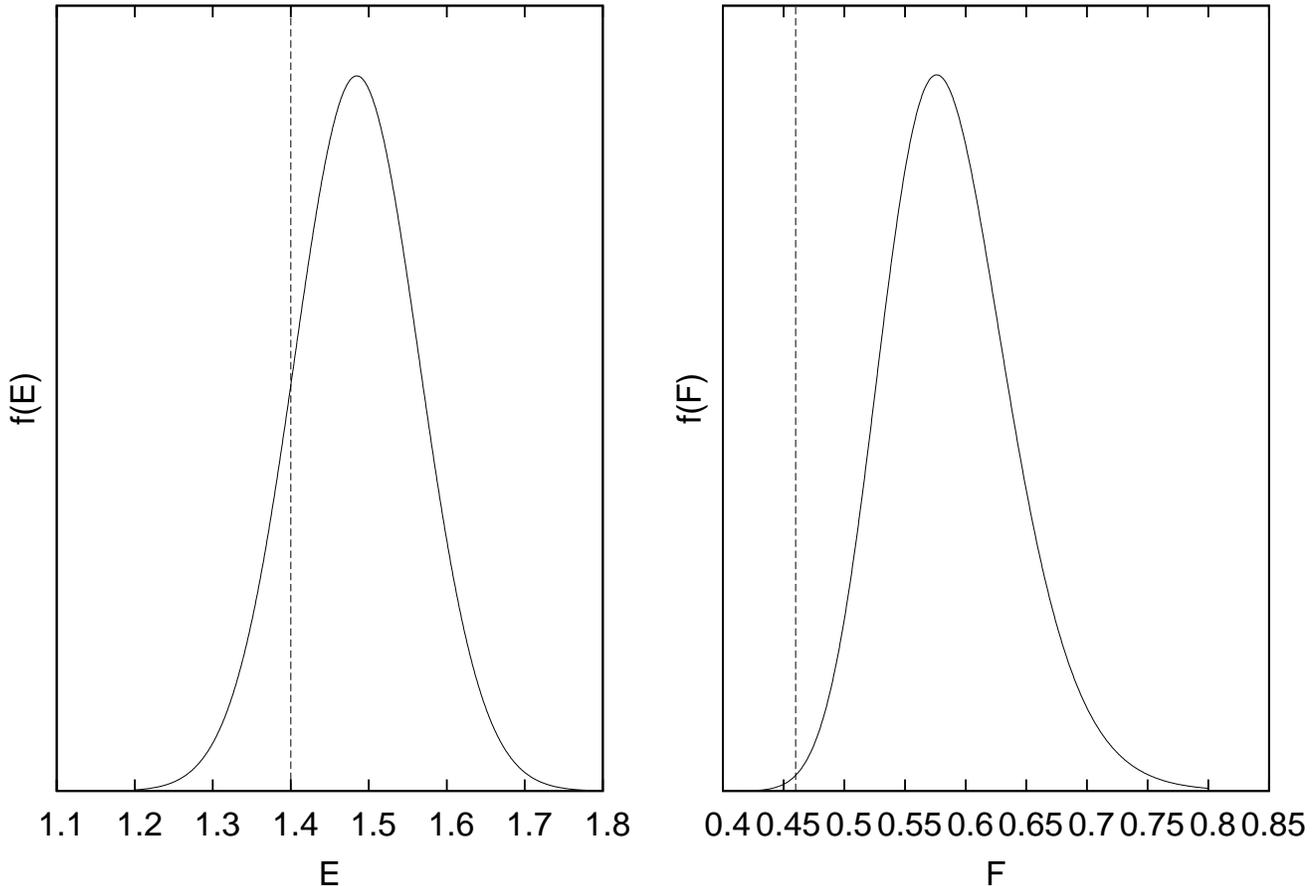,width=\textwidth,angle=270}} 
\caption{Marginalized PDFs for the log-normal parameters $E$ and $F$
  deduced from a fake population in which the true values were $E=1.4$
  and $F=0.46$ shown as dashed vertical lines. In this case, the
  assumed population parameters for the detectability analysis were
  intentionally biased to be $h=900$~pc, $\rho=4$, $\mu=-2.1$ and
  $\sigma=0.5$ and lead to a detectability model with $\alpha=2.2$ and
  $\beta=200$. Even with such a bias, the PDFs successfully encompass
  the true population values and favor the log-normal model by an
  order of magnitude over the three other models.}
\label{fig:fake}
\end{figure*}

While the above results are very encouraging, they represent idealized
conditions in which we input the actual values of $h$, $\rho$, $\mu$
and $\sigma$ into the detectability analysis to determine $\alpha$ and
$\beta$. In reality, of course, these numbers are not known and are
only approximations to the true distribution of MSPs. To examine how
robust the analysis is to changes in the assumed duty cycle, $h$,
$\rho$, $\mu$ and $\sigma$, we repeated the above procedure over a
range of values to determined $\alpha$ and $\beta$. The ranges we
explored were 5--30\% duty cycles, $300<h<900$~pc, $4<\rho<6$,
$-2.5<\mu<-1.5$ and $0.3<\sigma<1.5$. Although these led to variations
in the detectability parameters in the ranges $2 < \alpha < 15$~ms and
$100<\beta<300$~cm$^{-3}$~pc, we still found that the input parameter
distributions were recovered and that the correct distribution was
favored. An example of this is shown in Fig.~\ref{fig:fake} in which
we see the inferred PDFs from an analysis of a fake population with a
log-normal period distribution.  These results give us confidence that
our analysis on the observed sample of 56 MSPs is providing reliable
insights into their underlying spin period distribution, $f(P)$.

\subsection{Discussion}\label{sec:discuss}

Based on the analysis presented in this paper, we have found evidence
favoring the underlying spin period distribution of Galactic MSPs to
be log-normal in form. While a gamma distribution is compatible with
the data, it is less favoured than the log-normal. Uniform, power-law
and Gaussian distributions are decisively ruled out in our likelihood
analysis as being good descriptions to $f(P)$. We note that the strong
preference for a log-normal model found here is in contrast to the
power-law model proposed by \citet{cc97} based on a much smaller
sample of MSPs. While the exponent of our power-law model tested here
(--1.7) is consistent with theirs, the likelihood analysis strongly
favors the log-normal model.

\begin{figure} 
\centerline{\psfig{file=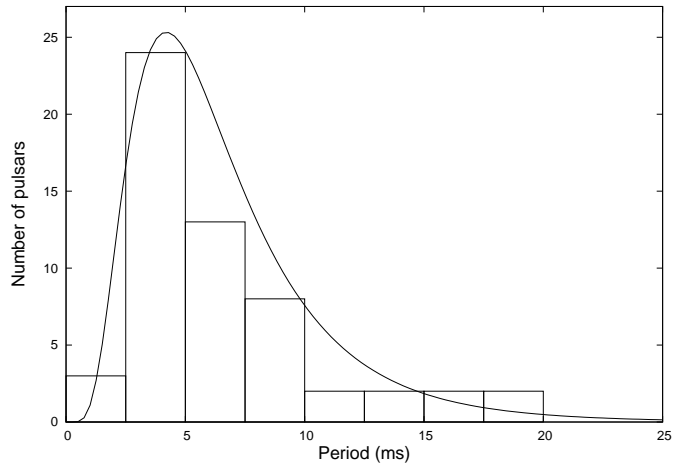,width=9cm,angle=270}} 
\caption{A comparison of the sample of 56 MSPs considered in this
  paper with our best-fitting period distribution from Equation~17
  (solid line).}
\label{fig:chi2}
\end{figure}

While our likelihood analysis weighs the different distributions we
tested against each other, some measure of the absolute agreement
between the log-normal model and the observed sample of 56 MSPs can be
found by comparing the sample with the predicted observed period
distribution for this model.  Combining our detectability model and
log-normal period distribution, the observed period distribution takes
the form
\begin{equation}
f_{\rm obs}(P) \propto \frac{1}{P} \exp \left[\frac{-(\ln(P)-E)^2}{2F^2}\right]
        \left(1 - \exp \left[ \frac{-P}{\alpha} \right] \right),
\end{equation}
where the log-normal parameters $E=1.5$ and $F=0.58$ and the
detectability parameter $\alpha=10$~ms. As can be seen from the
comparison of this function with the binned data from the 56-MSP
sample in Fig.~\ref{fig:chi2}, the agreement is excellent, with the
reduced $\chi^2$ value being 1.1.

\begin{figure} 
\centerline{\psfig{file=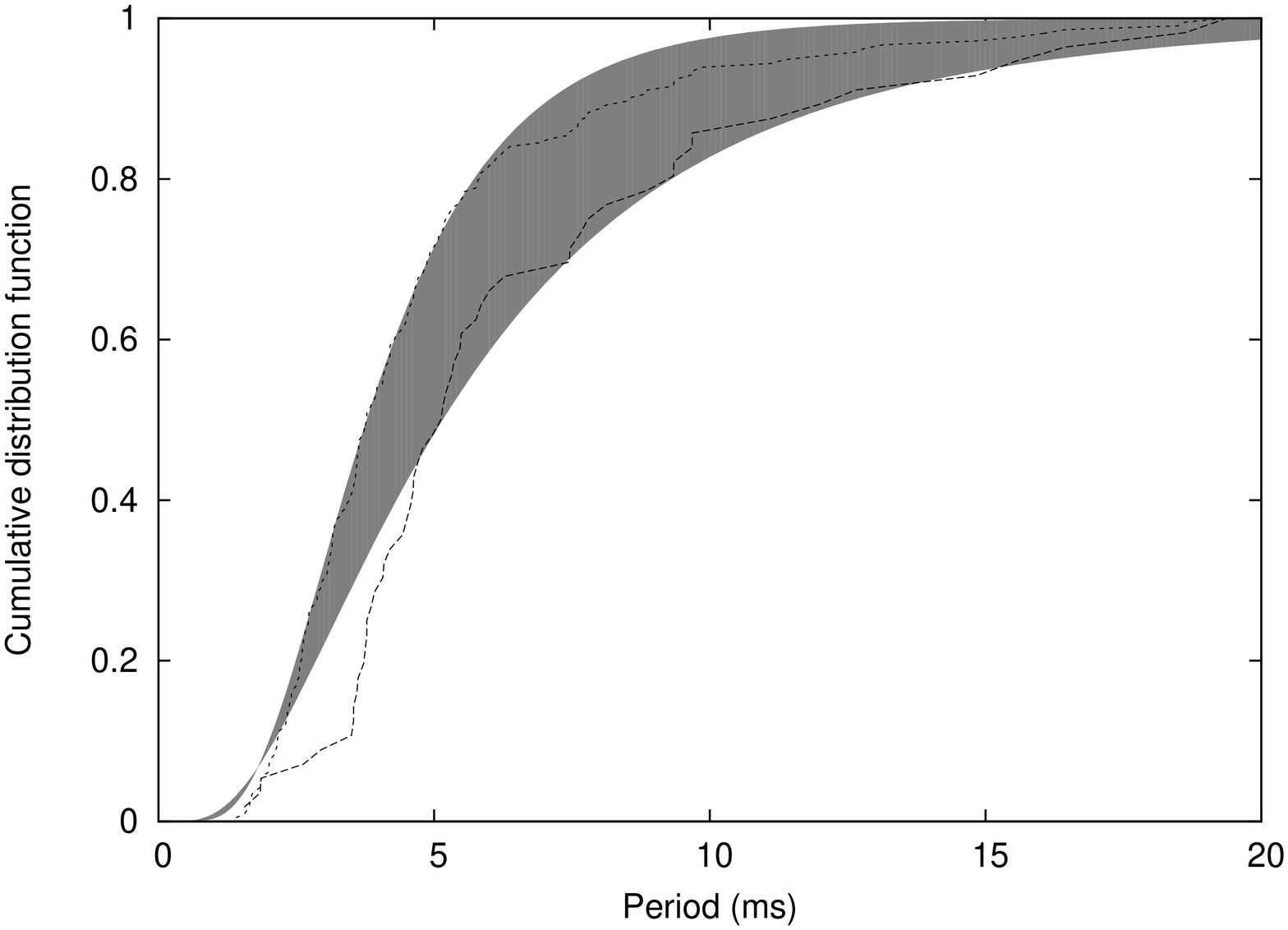,width=9.5cm,angle=270}} 
\caption{Cumulative distribution functions showing the observed sample
  of 56 MSPs (rightmost dashed curve), the 95\% credible region of our
  best fitting log-normal model (shaded band) and the current sample
  of 206 MSPs with $P<20$~ms (leftmost dashed curve). The deviation
  between the shaded band and the rightmost dashed curve highlights
  the observational selection effects at short periods in our sample
  of 56 MSPs.}
\label{fig:cdfs}
\end{figure}

Since the sample of MSPs used in this analysis is based on surveys
carried out a decade ago, it is useful to confront the distribution we
obtained with the present sample of objects. This is shown in
cumulative form in Fig.~\ref{fig:cdfs} where it is seen that the 95\%
credible region of log-normal functions we derive is broadly
compatible with the present sample of 228 MSPs which have been
detected in the Parkes High Time Resolution Universe Surveys
\citep{kjs+10,bck+13}, targeted searches of {\it Fermi} sources
\citep{rap+12} and also in surveys at lower frequencies with Arecibo
and Green Bank \citep{dsm+13,slr+14}. We note that the observed sample
lies to the upper end of the 95\% credible region shown in
Fig.~\ref{fig:cdfs}.  Future studies of this newer larger sample of
MSPs should, therefore, provide more stringent constraints on the
period distribution.

The general agreement with our log-normal model and the present sample
of MSPs suggests that the period-dependent selection effects on these
``first generation'' Parkes multibeam surveys (i.e.,~PMPS, PM, PA,
SWIL, SWHL and DMB) which we model in our detectability function are
much less severe in the present generation of MSP surveys.

\section{Conclusions}\label{sec:conc}

We have presented timing models for four MSPs found as part of the
Parkes Multibeam Pulsar survey of the Galactic plane.  From a
likelihood analysis of the sample of 56 MSPs detected with this
earlier generation of Parkes multibeam pulsar surveys, we demonstrate
that the underlying population of spin periods for MSPs is compatible
with a log-normal distribution. When this distribution is confronted
with more recent discoveries from other surveys, we see that it is
broadly consistent with the new results.  It is important to note that
the distribution we have derived here applies to the present-day MSP
population. Although to first order, because of the very low spin-down
rates of MSPs, the birth spin-period distribution may not be
significantly different \citep[see, e.g.,][]{ctk94}, further
investigations are necessary to confirm this conjecture.

Although the true period distribution for MSPs may not be as simple as
our analysis might initially suggest, it is clear that the
distributions considered here all allow for the existence of a small
fraction of pulsars with $P<1.5$~ms. Based on the smooth curves shown
in Fig.~\ref{fig:cdfs}, the fraction of such pulsars in the population
is around 3\%.  The true fraction could even be higher than this if we
have overestimated the detectability of such rapidly spinning pulsars.
Given our estimate of the analytic form of the observed period
distribution given in Equation 17, we find that the probability of not
detecting a pulsar in our sample of 56 MSPs with period $P<1.5$~ms in
the current sample is 99.2\%. This is entirely consistent with the
lack of such pulsars in the sample so far.  Further discussion about
the possibility of sub-millisecond pulsars can be found in
\citet{lbb+13} and references therein.

An inspection of the current sample of MSPs shows no statistically
significant difference between the spin period distributions of
isolated objects versus binary systems.  A useful approach which is
currently being pursued (P.~Lazarus, private communication) is to
artificially add short period signals to existing search data sets and
directly test the effectiveness of pulsar search codes in recovering
these signals.  The modeling techniques presented here may be useful
in further analyses of the MSP population which need to take account
of the different selection biases and observing frequencies that have
taken place since the completion of the initial Parkes multibeam
surveys.  The techniques may also be applied to other population
parameters in which selection effects may be apparent, for example the
$P-\dot{P}$ distribution.

\section*{Acknowledgments} 

The Parkes radio telescope is part of the Australia Telescope which is
funded by the Commonwealth of Australia for operation as a National
Facility managed by CSIRO.  DRL acknowledges support from the Royal
Society as a University Research Fellow during the early phases of
this project.  DRL and MAM acknowledge support from Oxford
Astrophysics while on sabbatical leave in 2013.  PE acknowledges a
Fulbright Research Scholar grant administered by the U.S.--Italy
Fulbright Commission and is grateful to the Harvard--Smithsonian
Center for Astrophysics for hosting him during his Fulbright
exchange. The Fulbright Scholar Program is sponsored by the
U.S. Department of State and administered by CIES, a division of
IIE. Pulsar Research at UBC is supported by an NSERC Discovery
Grant. Current support to DRL and MAM is provided by the National
Science Foundation PIRE award 0968296.  DRL thanks Joris Verbiest for
useful discussions and to Simon Johnston and the CSIRO staff for their
hospitality during the final stages of this work. We thank the referee
for useful comments on the originally submitted version of this paper.


\begin{thebibliography}{50}
\expandafter\ifx\csname natexlab\endcsname\relax\def\natexlab#1{#1}\fi

\bibitem[{Backer {et~al}\mbox{.}(1982)Backer, Kulkarni, Heiles, Davis, \&
  Goss}]{bkh+82}
Backer D.~C., Kulkarni S.~R., Heiles C., Davis M.~M., Goss W.~M., 1982, Nature,
  300, 615

\bibitem[{{Bagchi} {et~al}\mbox{.}(2011){Bagchi}, {Lorimer}, \&
  {Chennamangalam}}]{blc11}
{Bagchi} M., {Lorimer} D.~R., {Chennamangalam} J., 2011, MNRAS, 418, 477

\bibitem[{{Barr} {et~al}\mbox{.}(2013){Barr}, {Champion}, {Kramer}, {Eatough},
  {Freire}, {Karuppusamy}, {Lee}, {Verbiest}, {Bassa}, {Lyne}, {Stappers},
  {Lorimer}, \& {Klein}}]{bck+13}
{Barr} E.~D. {et~al.}, 2013, MNRAS, 435, 2234

\bibitem[{Blandford \& Teukolsky(1976)}]{bt76}
Blandford R., Teukolsky S.~A., 1976, ApJ, 205, 580

\bibitem[{{Burgay} {et~al}\mbox{.}(2006){Burgay}, {Joshi}, {D'Amico},
  {Possenti}, {Lyne}, {Manchester}, {McLaughlin}, {Kramer}, {Camilo}, \&
  {Freire}}]{bjd+06}
{Burgay} M. {et~al.}, 2006, MNRAS, 368, 283

\bibitem[{{Burgay} {et~al}\mbox{.}(2013){Burgay}, {Keith}, {Lorimer},
  {Hassall}, {Lyne}, {Camilo}, {D'Amico}, {Hobbs}, {Kramer}, {Manchester},
  {McLaughlin}, {Possenti}, {Stairs}, \& {Stappers}}]{bkl+13}
{Burgay} M. {et~al.}, 2013, MNRAS, 429, 579

\bibitem[{Camilo {et~al}\mbox{.}(2001)Camilo, Lyne, Manchester, Bell, Stairs,
  D'Amico, Kaspi, Possenti, Crawford, \& McKay}]{clm+01}
Camilo F. {et~al.}, 2001, ApJ, 548, L187

\bibitem[{Camilo {et~al}\mbox{.}(1994)Camilo, Thorsett, \& Kulkarni}]{ctk94}
Camilo F., Thorsett S.~E., Kulkarni S.~R., 1994, ApJ, 421, L15

\bibitem[{Cordes \& Chernoff(1997)}]{cc97}
Cordes J.~M., Chernoff D.~F., 1997, ApJ, 482, 971

\bibitem[{{Cordes} \& {Lazio}(2002)}]{cl02}
{Cordes} J.~M., {Lazio} T.~J.~W., 2002, preprint (astro-ph/0207156)

\bibitem[{{Crawford} {et~al}\mbox{.}(2013){Crawford}, {Lyne}, {Stairs},
  {Kaplan}, {McLaughlin}, {Freire}, {Burgay}, {Camilo}, {D'Amico}, {Faulkner},
  {Kramer}, {Lorimer}, {Manchester}, {Possenti}, \& {Steeghs}}]{cls+13}
{Crawford} F. {et~al.}, 2013, ApJ, 776, 20

\bibitem[{Damour \& Taylor(1991)}]{dt91}
Damour T., Taylor J.~H., 1991, ApJ, 366, 501

\bibitem[{{Deneva} {et~al}\mbox{.}(2013){Deneva}, {Stovall}, {McLaughlin},
  {Bates}, {Freire}, {Martinez}, {Jenet}, \& {Bagchi}}]{dsm+13}
{Deneva} J.~S., {Stovall} K., {McLaughlin} M.~A., {Bates} S.~D., {Freire}
  P.~C.~C., {Martinez} J.~G., {Jenet} F., {Bagchi} M., 2013, ApJ, 775, 51

\bibitem[{{Eatough} {et~al}\mbox{.}(2009){Eatough}, {Keane}, \& {Lyne}}]{ekl09}
{Eatough} R.~P., {Keane} E.~F., {Lyne} A.~G., 2009, MNRAS, 395, 410

\bibitem[{{Eatough} {et~al}\mbox{.}(2013){Eatough}, {Kramer}, {Lyne}, \&
  {Keith}}]{eklk13}
{Eatough} R.~P., {Kramer} M., {Lyne} A.~G., {Keith} M.~J., 2013, MNRAS, 431,
  292

\bibitem[{{Edwards} {et~al}\mbox{.}(2001){Edwards}, {Bailes}, {van Straten}, \&
  {Britton}}]{ebvb01}
{Edwards} R.~T., {Bailes} M., {van Straten} W., {Britton} M.~C., 2001, MNRAS,
  326, 358

\bibitem[{{Faucher-Gigu{\`e}re} \& {Kaspi}(2006)}]{fk06}
{Faucher-Gigu{\`e}re} C.-A., {Kaspi} V.~M., 2006, ApJ, 643, 332

\bibitem[{{Faulkner} {et~al}\mbox{.}(2004){Faulkner}, {Stairs}, {Kramer},
  {Lyne}, {Hobbs}, {Possenti}, {Lorimer}, {Manchester}, {McLaughlin},
  {D'Amico}, {Camilo}, \& {Burgay}}]{fsk+04}
{Faulkner} A.~J. {et~al.}, 2004, MNRAS, 355, 147

\bibitem[{{Folkner} {et~al}\mbox{.}(2008){Folkner}, {Williams}, \&
  {Boggs}}]{fwb08}
{Folkner} W.~M., {Williams} J.~G., {Boggs} D.~H., 2008, {JPL Memo series},
  343R-08-003

\bibitem[{{Hobbs} {et~al}\mbox{.}(2010){Hobbs}, {Archibald}, {Arzoumanian},
  {Backer}, {Bailes}, {Bhat}, {Burgay}, {Burke-Spolaor}, {Champion}, {Cognard},
  {Coles}, {Cordes}, {Demorest}, {Desvignes}, {Ferdman}, {Finn}, {Freire},
  {Gonzalez}, {Hessels}, {Hotan}, {Janssen}, {Jenet}, {Jessner}, {Jordan},
  {Kaspi}, {Kramer}, {Kondratiev}, {Lazio}, {Lazaridis}, {Lee}, {Levin},
  {Lommen}, {Lorimer}, {Lynch}, {Lyne}, {Manchester}, {McLaughlin}, {Nice},
  {Oslowski}, {Pilia}, {Possenti}, {Purver}, {Ransom}, {Reynolds}, {Sanidas},
  {Sarkissian}, {Sesana}, {Shannon}, {Siemens}, {Stairs}, {Stappers},
  {Stinebring}, {Theureau}, {van Haasteren}, {van Straten}, {Verbiest},
  {Yardley}, \& {You}}]{haa+10}
{Hobbs} G. {et~al.}, 2010, Classical and Quantum Gravity, 27, 084013

\bibitem[{{Hobbs} {et~al}\mbox{.}(2004){Hobbs}, {Faulkner}, {Stairs}, {Camilo},
  {Manchester}, {Lyne}, {Kramer}, {D'Amico}, {Kaspi}, {Possenti}, {McLaughlin},
  {Lorimer}, {Burgay}, {Joshi}, \& {Crawford}}]{hfs+04}
{Hobbs} G. {et~al.}, 2004, MNRAS, 352, 1439

\bibitem[{{Hobbs} {et~al}\mbox{.}(2006){Hobbs}, {Edwards}, \&
  {Manchester}}]{hem06}
{Hobbs} G.~B., {Edwards} R.~T., {Manchester} R.~N., 2006, MNRAS, 369, 655

\bibitem[{{Hotan} {et~al}\mbox{.}(2004){Hotan}, {van Straten}, \&
  {Manchester}}]{hvm04}
{Hotan} A.~W., {van Straten} W., {Manchester} R.~N., 2004, Publ. Astron. Soc. Aust., 21, 302

\bibitem[{{Jacoby} {et~al}\mbox{.}(2007){Jacoby}, {Bailes}, {Ord}, {Knight}, \&
  {Hotan}}]{jbo+07}
{Jacoby} B.~A., {Bailes} M., {Ord} S.~M., {Knight} H.~S., {Hotan} A.~W., 2007,
  ApJ, 656, 408

\bibitem[{Jeffreys(1961)}]{jef61}
Jeffreys H., 1961, {Theory of Probability}. Oxford University Press

\bibitem[{Johnston \& Bailes(1991)}]{jb91}
Johnston S., Bailes M., 1991, MNRAS, 252, 277

\bibitem[{{Keane} {et~al}\mbox{.}(2010){Keane}, {Ludovici}, {Eatough},
  {Kramer}, {Lyne}, {McLaughlin}, \& {Stappers}}]{kle+10}
{Keane} E.~F., {Ludovici} D.~A., {Eatough} R.~P., {Kramer} M., {Lyne} A.~G.,
  {McLaughlin} M.~A., {Stappers} B.~W., 2010, MNRAS, 401, 1057

\bibitem[{{Keane} {et~al}\mbox{.}(2012){Keane}, {Stappers}, {Kramer}, \&
  {Lyne}}]{kskl12}
{Keane} E.~F., {Stappers} B.~W., {Kramer} M., {Lyne} A.~G., 2012, MNRAS, 425,
  L71

\bibitem[{{Keith} {et~al}\mbox{.}(2009){Keith}, {Eatough}, {Lyne}, {Kramer},
  {Possenti}, {Camilo}, \& {Manchester}}]{kel+09}
{Keith} M.~J., {Eatough} R.~P., {Lyne} A.~G., {Kramer} M., {Possenti} A.,
  {Camilo} F., {Manchester} R.~N., 2009, MNRAS, 395, 837

\bibitem[{{Keith} {et~al}\mbox{.}(2010){Keith}, {Jameson}, {van Straten},
  {Bailes}, {Johnston}, {Kramer}, {Possenti}, {Bates}, {Bhat}, {Burgay},
  {Burke-Spolaor}, {D'Amico}, {Levin}, {McMahon}, {Milia}, \&
  {Stappers}}]{kjs+10}
{Keith} M.~J. {et~al.}, 2010, MNRAS, 409, 619

\bibitem[{{Knispel} {et~al}\mbox{.}(2013){Knispel}, {Eatough}, {Kim}, {Keane},
  {Allen}, {Anderson}, {Aulbert}, {Bock}, {Crawford}, {Eggenstein}, {Fehrmann},
  {Hammer}, {Kramer}, {Lyne}, {Machenschalk}, {Miller}, {Papa}, {Rastawicki},
  {Sarkissian}, {Siemens}, \& {Stappers}}]{kek+13}
{Knispel} B. {et~al.}, 2013, ApJ, 774, 93

\bibitem[{{Kramer} {et~al}\mbox{.}(2003){Kramer}, {Bell}, {Manchester}, {Lyne},
  {Camilo}, {Stairs}, {D'Amico}, {Kaspi}, {Hobbs}, {Morris}, {Crawford},
  {Possenti}, {Joshi}, {McLaughlin}, {Lorimer}, \& {Faulkner}}]{kbm+03}
{Kramer} M. {et~al.}, 2003, MNRAS, 342, 1299

\bibitem[{Kulkarni \& Narayan(1988)}]{kn88}
Kulkarni S.~R., Narayan R., 1988, ApJ, 335, 755

\bibitem[{Lange {et~al}\mbox{.}(2001)Lange, Camilo, Wex, Kramer, Backer, Lyne,
  \& Doroshenko}]{lcw+01}
Lange C., Camilo F., Wex N., Kramer M., Backer D., Lyne A., Doroshenko O.,
  2001, MNRAS, 326, 274

\bibitem[{{Levin} {et~al}\mbox{.}(2013){Levin}, {Bailes}, {Barsdell}, {Bates},
  {Bhat}, {Burgay}, {Burke-Spolaor}, {Champion}, {Coster}, {D'Amico},
  {Jameson}, {Johnston}, {Keith}, {Kramer}, {Milia}, {Ng}, {Possenti},
  {Stappers}, {Thornton}, \& {van Straten}}]{lbb+13}
{Levin} L. {et~al.}, 2013, MNRAS, 434, 1387

\bibitem[{{Lorimer}(2013)}]{lor13}
{Lorimer} D.~R., 2013, in IAU Symposium, Vol. 291, IAU Symposium, {van Leeuwen}
  J., ed., pp. 237--242

\bibitem[{Lorimer {et~al}\mbox{.}(2013)Lorimer, Camilo, \& McLaughlin}]{lcm13}
Lorimer D.~R., Camilo F., McLaughlin M.~A., 2013, MNRAS, 434, 50

\bibitem[{{Lorimer} {et~al}\mbox{.}(2006){Lorimer}, {Faulkner}, {Lyne},
  {Manchester}, {Kramer}, {McLaughlin}, {Hobbs}, {Possenti}, {Stairs},
  {Camilo}, {Burgay}, {D'Amico}, {Corongiu}, \& {Crawford}}]{lfl+06}
{Lorimer} D.~R. {et~al.}, 2006, MNRAS, 372, 777

\bibitem[{Lorimer \& Kramer(2005)}]{lk05}
Lorimer D.~R., Kramer M., 2005, {Handbook of Pulsar Astronomy}. Cambridge
  University Press

\bibitem[{{Manchester} {et~al}\mbox{.}(2013){Manchester}, {Hobbs}, {Bailes},
  {Coles}, {van Straten}, {Keith}, {Shannon}, {Bhat}, {Brown}, {Burke-Spolaor},
  {Champion}, {Chaudhary}, {Edwards}, {Hampson}, {Hotan}, {Jameson}, {Jenet},
  {Kesteven}, {Khoo}, {Kocz}, {Maciesiak}, {Oslowski}, {Ravi}, {Reynolds},
  {Sarkissian}, {Verbiest}, {Wen}, {Wilson}, {Yardley}, {Yan}, \&
  {You}}]{mhb+13}
{Manchester} R.~N. {et~al.}, 2013, Publ. Astron. Soc. Aust., 30, 17

\bibitem[{Manchester {et~al}\mbox{.}(2001)Manchester, Lyne, Camilo, Bell,
  Kaspi, D'Amico, McKay, Crawford, Stairs, Possenti, Morris, \&
  Sheppard}]{mlc+01}
Manchester R.~N. {et~al.}, 2001, MNRAS, 328, 17

\bibitem[{McLaughlin {et~al}\mbox{.}(2006)McLaughlin, Lyne, Lorimer, Kramer,
  Faulkner, Manchester, Cordes, Possenti, Camilo, Hobbs, Stairs, D'Amico, \&
  O'Brien}]{mll+06}
McLaughlin M.~A. {et~al.}, 2006, Nature, 439, 817

\bibitem[{{Mickaliger} {et~al}\mbox{.}(2012){Mickaliger}, {Lorimer}, {Boyles},
  {McLaughlin}, {Collins}, {Hough}, {Tehrani}, {Tenney}, {Liska}, \&
  {Swiggum}}]{mlb+12}
{Mickaliger} M.~B. {et~al.}, 2012, ApJ, 759, 127

\bibitem[{{Morris} {et~al}\mbox{.}(2002){Morris}, {Hobbs}, {Lyne}, {Stairs},
  {Camilo}, {Manchester}, {Possenti}, {Bell}, {Kaspi}, {Amico}, {McKay},
  {Crawford}, \& {Kramer}}]{mhl+02}
{Morris} D.~J. {et~al.}, 2002, MNRAS, 335, 275

\bibitem[{Nice \& Taylor(1995)}]{nt95}
Nice D.~J., Taylor J.~H., 1995, ApJ, 441, 429

\bibitem[{{Ray} {et~al}\mbox{.}(2012){Ray}, {Abdo}, {Parent}, {Bhattacharya},
  {Bhattacharyya}, {Camilo}, {Cognard}, {Theureau}, {Ferrara}, {Harding},
  {Thompson}, {Freire}, {Guillemot}, {Gupta}, {Roy}, {Hessels}, {Johnston},
  {Keith}, {Shannon}, {Kerr}, {Michelson}, {Romani}, {Kramer}, {McLaughlin},
  {Ransom}, {Roberts}, {Saz Parkinson}, {Ziegler}, {Smith}, {Stappers},
  {Weltevrede}, \& {Wood}}]{rap+12}
{Ray} P.~S. {et~al.}, 2012, Fermi Symposium proceedings - eConf C110509
(arXiv:1205.3089)

\bibitem[{Shklovskii(1970)}]{shk70}
Shklovskii I.~S., 1970, Sov. Astron., 13, 562

\bibitem[{{Stovall} {et~al}\mbox{.}(2013){Stovall}, {Lorimer}, \&
  {Lynch}}]{sll13}
{Stovall} K., {Lorimer} D.~R., {Lynch} R.~S., 2013, Classical and Quantum
  Gravity, 30, 224003

\bibitem[{{Stovall} {et~al}\mbox{.}(2014){Stovall}, {Lynch}, {Ransom},
  {Archibald}, {Banaszak}, {Biwer}, {Boyles}, {Dartez}, {Day}, {Ford},
  {Flanigan}, {Garcia}, {Hessels}, {Hinojosa}, {Jenet}, {Kaplan},
  {Karako-Argaman}, {Kaspi}, {Kondratiev}, {Leake}, {Lorimer}, {Lunsford},
  {Martinez}, {Mata}, {McLaughlin}, {Roberts}, {Rohr}, {Siemens}, {Stairs},
  {van Leeuwen}, {Walker}, \& {Wells}}]{slr+14}
{Stovall} K. {et~al.}, 2014, ApJ, 791, 67

\bibitem[{{Verbiest} {et~al}\mbox{.}(2012){Verbiest}, {Weisberg}, {Chael},
  {Lee}, \& {Lorimer}}]{vwc+12}
{Verbiest} J.~P.~W., {Weisberg} J.~M., {Chael} A.~A., {Lee} K.~J., {Lorimer}
  D.~R., 2012, ApJ, 755, 39

\end{thebibliography}
\end{document}